\documentclass[onecolumn,aps,prc,nofootinbib]{revtex4}
\usepackage{amsfonts}
\usepackage{epsfig}
\usepackage{epsfig}
\usepackage{epsfig}
\usepackage{amsmath}
\usepackage{bm}
\usepackage{graphicx}

\newcommand{\beq}{\begin{equation}}
\newcommand{\eeq}{\end{equation}}
\newcommand{\bea}{\vspace{0.25cm}\begin{eqnarray}}
\newcommand{\eea}{\end{eqnarray}}

\def\lsim{\mathrel{\rlap{\lower4pt\hbox{\hskip1pt$\sim$}}
    \raise1pt\hbox{$<$}}}         
\def\gsim{\mathrel{\rlap{\lower4pt\hbox{\hskip1pt$\sim$}}
    \raise1pt\hbox{$>$}}}         

\long\def\symbolfootnote[#1]#2{\begingroup%
\def\thefootnote{\fnsymbol{footnote}}\footnotemark[#1]\footnotetext[#1]{#2}\endgroup}

\begin{document}

\title{
Jet quenching in mini-quark-gluon plasma:
  Medium modification factor $I_{pA}$ for photon-tagged jets
}

\author{B.G. Zakharov}

\affiliation{
L.D.~Landau Institute for Theoretical Physics,
        GSP-1, 117940,\\ Kosygina Str. 2, 117334 Moscow, Russia
}

\begin{abstract}
  We calculate the medium modification factor $I_{pA}$
  for the photon-tagged jet fragmentation functions
  for scenario with the quark-gluon plasma formation
  in $pA$ and $pp$ collisions.
  We perform calculations of radiative and collisional parton
  energy loss in the quark-gluon plasma
  with running $\alpha_s$ which has a plateau around $Q\sim \kappa T$
  with $\kappa$ fitted to the LHC data on the heavy ion $R_{AA}$.
We find that the theoretical prediction for $I_{pA}$
in $5.02$ TeV $p$+Pb collisions
are within errors consistent with the data from ALICE \cite{ALICE_IpA}.
However, a definite conclusion about the presence or absence of jet quenching
in $pA$ collisions
cannot be drawn due to large experimental errors of the ALICE data
\cite{ALICE_IpA}.
Our calculations show that this requires a significantly more accurate
measurement of $I_{pA}$.

\end{abstract}
%

\maketitle

{\bf Introduction}.
Heavy ion collision experiments at RHIC and the LHC led to
the discovery of the Quark Gluon Plasma (QGP) formation
in $AA$ collisions. The most striking manifestations
of the QGP formation in $AA$ collisions are
the transverse flow effects in the azimuthal correlations for soft
hadrons and the strong suppression
of high-$p_T$ hadron spectra (jet quenching). 
Hydrodynamic analyses of soft hadron production in $AA$
collisions show that the QCD matter produced in $AA$ collisions
flows almost as a perfect fluid (for reviews, see, e.g., Refs.
\cite{hydro2,hydro3}).
Jet quenching in $AA$ collisions is due to radiative
and collisional energy loss of fast partons in the hot QGP.
The dominant contribution to the parton energy loss comes from the radiative
mechanism due to induced gluon radiation \cite{GW,BDMPS1,LCPI1,W1,GLV1,AMY1}.
The available data from RHIC and the LHC on the nuclear modification
factor $R_{AA}$ of hadron spectra in $AA$ collisions can be described
in the pQCD picture of parton energy loss for the QGP formation time
$\tau_0\sim 0.5-1$ fm
(see, e.g., Refs. \cite{CUJET3,Armesto_LHC,Z_hl})
that is roughly consistent with the results of hydrodynamical analyses
of experimental data on $AA$ collisions \cite{Heinz_tau}.

In recent years, the azimuthal correlations in soft hadron production
(the ridge effect), similar to that observed in $AA$ collisions,
have been observed in $pp$/$pA$ collisions.   
The formation of a mini QGP (mQGP) fireball is the most popular explanation
of the ridge/flow effects in $pp$/$pA$ collisions (for a review, see
Ref. \cite{Schenke_mQGP}).
There are several experimental evidences supporting
the onset of the mQGP regime in $pp$/$pA$ collisions at the charged hadron
multiplicity density $dN_{ch}/d\eta\gsim 5$
\cite{ALICE_strange,Camp1}.
It is important that, from the point of view of the multiplicity
density, conditions for the mQGP formation
in $pp$/$pA$ collisions are more favorable for events with jet production.
Because in jet events the average multiplicity density
of soft (underlying-event (UE)) hadrons
is larger than the minimum-bias multiplicity by a factor of
$\sim 2-2.5$ \cite{Field}.
At the LHC energies in $pp$ jet events we have 
 $dN_{ch}^{ue}/d\eta\sim 10-15$   (and by a factor of $\sim 2-3$ larger values
for $pA$ collisions), that seems to be large enough to expect
the mQGP formation (in the light of the results of
\cite{ALICE_strange,Camp1}).
In the scenario with the mQGP formation in $pp/pA$ collisions,
the jet quenching effects must appear.
Similarly to $AA$ collisions, they should modify
the jet fragmentation functions (FFs) and hadron spectra in $pp$/$pA$
collisions as compared to predictions of the standard pQCD.
The recent ALICE \cite{ALICE_Ipp} measurement
of the jet FF modification factor $I_{pp}$
for the hadron-tagged jets in $pp$ collisions at $\sqrt{s}=5.02$ TeV
 seems to confirm the scenario
with the mQGP formation and jet quenching in $pp$ collisions,
since the data \cite{ALICE_Ipp} show a monotonic decrease of $I_{pp}$
with the UE multiplicity expected for the scenario with the mQGP formation
\cite{Z_pp_PRL}. 
The results of \cite{ALICE_Ipp} agree within errors
with calculations of \cite{Z_Ipp} in the framework of the
light-cone path integral (LCPI) approach to induced gluon emission
\cite{LCPI1}.

The first calculations of the medium modification factor $R_{pp}$ for
$pp$ collisions were performed in \cite{Z_pp_PRL,Z_pp13}
within the LCPI formalism  for induced gluon emission \cite{LCPI1}.
These calculations
(and the more recent and accurate analysis \cite{Z_hl})
show that $R_{pp}$ is close to unity. 
The $R_{pp}$ does not admit a direct measurement, but
it modifies a little theoretical predictions
for $R_{AA}$.
However, in \cite{Z_hl} it was demonstrated that the available data on $R_{AA}$
can be described fairly well both in the scenarios with and without the mQGP
formation in $pp$ collisions.
It is believed that measurement of the nuclear modification factor
$R_{pA}$ for high-$p_T$ hadrons in $pA$ collisions
is a promising method for observation of
jet quenching caused by the mQGP formation.
$R_{pA}$ is defined as the ratio of the $pA$ spectrum
to the binary scaled $pp$ one, and, contrary to $R_{pp}$,
it is a measurable quantity ($R_{pA}\ne 1$ even without the final state
  interaction effects,  due to the difference between the nuclear
parton distribution functions  (PDFs) and the proton PDFs  
 (which we denote by $R_{pA}^{PDF}$)).
It is reasonable to expect that for the scenario with the mQGP formation
both in $pA$ and $pp$ collisions, jet quenching should
be stronger in $pA$ collisions,
as a result the experimental
$R_{pA}^{exp}$ should be smaller than $R_{pA}^{PDF}$.
The available
experimental data \cite{CMS_RPA5,ALICE_RPA5,ALICE_RPA8} on $R_{pA}$
are controversial: there is a significant discrepancy
between data from CMS \cite{CMS_RPA5} for 5.02 TeV $p$+Pb collisions
($R_{pPb}\sim 1.1-1.19$ at $p_T\gsim 10$ TeV)
and data from ALICE \cite{ALICE_RPA5,ALICE_RPA8} for 5.02 and 8.16 $p$+Pb
collision ($R_{pPb}\sim 0.9-1.1$).
Calculations of \cite{Z_hl,Z_RpA} show that the data from CMS \cite{CMS_RPA5}
are clearly inconsistent with the scenario
with the mQGP formation,
but the data on $R_{pPb}$ from ALICE \cite{ALICE_RPA5,ALICE_RPA8}
may be consistent with the mQGP formation (both in $pp$ and $p$+Pb collisions).

Another way to probe the jet quenching effects in $pA$
collisions is measurement of the medium modification factor $I_{pA}$ for
the photon-tagged FFs for $\gamma$+jet events.
In analogy with
the medium modification factor $I_{AA}$ in $AA$ collisions (see e.g.
\cite{Wang1,STAR1}), $I_{pA}$, for a given photon transverse momentum
$p_T^{\gamma}$, is defined as the ratio 
\beq
I_{pA}(z_T,p_T^{\gamma})=D_h^{pA}(z_T,p_T^{\gamma})/D_h^{pp}(z_T,p_T^{\gamma})\,,
\label{eq:10}
\eeq
where $D_h^{pA,pp}$ are the photon-tagged FFs
of the away-side hard partons to the associate charged hadron $h$
for $pA$ and $pp$ collisions, $z_T=p_T^h/p_T^{\gamma}$, and $p_T^h$
is the hadron transverse momentum.
Experimentally, the photon-tagged FF $D_h$ is the away-side associated hadron
yield per trigger photon.
In terms of the inclusive cross sections, $D_h$ reads
\beq
D_h(z_T,p_T^{\gamma})=
\frac{p_T^{\gamma}d^3\sigma}{dp_T^hdp_T^{\gamma}dy^{\gamma}}
\Big(\frac{d^2\sigma}{dp_T^{\gamma}dy^{\gamma}}\Big)^{-1}\,.
\label{eq:20}
\eeq
The advantage of $I_{pA}$ is that experimental $D_h$ do not suffer
from the uncertainties of the yield normalizations in $pA$/$pp$ collisions
(since both the numerator and the denominator in (\ref{eq:20})
are hard cross
sections, and the normalization uncertainties
are largely canceled in $D_h$).
For the same reason, the theoretical $I_{pA}$, contrary to $R_{pA}$, is
insensitive to uncertainties in the nuclear and proton PDFs.

Recently,
the midrapidity $I_{pA}$
has been measured by the ALICE collaboration \cite{ALICE_IpA}
for 5.02 TeV $p$+Pb collisions
for the trigger photon momentum $12<p_T^{\gamma}<40$ GeV.
The ALICE measurement gives
$\langle I_{pA}\rangle\approx 0.84\pm 0.11$(stat)$\pm 0.19$(sys).
The $z_T$-dependence of $I_{pA}$ obtained in \cite{ALICE_IpA}
has some tendency of $I_{pA}$ towards decrease with increasing $z_T$.
This pattern, at least roughly, is what is expected in the scenario
with the mQGP formation. 
Of course, to understand better whether the results of \cite{ALICE_IpA}
are consistent with the scenario with the mQGP formation in $pp$/$pA$
collisions, quantitative calculations 
of $I_{pA}$ for this scenario are necessary.
In this paper, we perform calculations of $I_{pA}$ for conditions
of the ALICE experiment
\cite{ALICE_IpA}.
We use the LCPI approach \cite{LCPI1} to induced gluon emission
with temperature dependent $\alpha_s$ \cite{RAA20T},
which has successfully been used in our recent 
analysis \cite{Z_hl}  of the available data on the nuclear
modification factor $R_{AA}$.\\


{\bf Outline of the jet quenching scheme for FFs}.
We treat the $\gamma$+jet process in leading order (LO) pQCD.
In this approximation the transverse momentum of the hard parton,
produced in the direction opposite to the direct photon,
equals the photon transverse momentum.
The higher order effects lead to
fluctuation of the away side parton transverse momentum
around $p_T^{\gamma}$.
In \cite{Z_IAA17}, using the results of the
NLO pQCD analysis of the direct photon production of Ref.~\cite{Wang_NLO2},
it was demonstrated that for the trigger photon momentum
$p_T^{\gamma}\gsim 12$ GeV the smearing correction to the medium modification
factor $I_{AA}$ (which is
$\propto dI_{AA}/dz/p_{T}^{\gamma\,\,2}$)
is very small at $z_T\lsim 0.85-0.9$.
Since the magnitude of the jet modification
for $pA$ collisions is considerably smaller than that in $AA$ collisions,
the effect of smearing on $I_{pA}$ should also be smaller. This allows us
to ignore the smearing correction to $I_{pA}$ (except for $z_T$ very close to unity).

In the LO pQCD, when the photon transverse momentum coincides with
that for the away side parton,
the photon-tagged jet FFs defined by (\ref{eq:20})
can be decomposed as
\beq
D_{h}(z_T,p_{T}^{\gamma})=
\sum_{i} r_{i}(p_{T}^{\gamma})D_{h/i}(z_T,p_{T}^{i})\,,
\label{eq:30}
\eeq
where $D_{h/i}(z_T,p_T^i)$ is the FF for
transition of the initial parton $i$ with momentum $p_T^i=p_T^{\gamma}$
into the final hadron $h$, and 
$r_{i}$ is the relative weight of the $\gamma+i$ state in the
jet production. For scenario without the mQGP formation,
$D_{h/i}$ in (\ref{eq:30}) is the ordinary vacuum FF, $D_{h/i}^v$, and
for scenario with the mQGP formation $D_{h/i}$ is the medium modified
FF, $D_{h/i}^{m}$, (averaged over the jet path length $L$ in the mQGP).
We calculate the hard parton cross sections 
with the CTEQ6 \cite{CTEQ6}  PDFs
(with the EPS09 correction \cite{EPS09} for the nuclear PDFs).
For 5.02 TeV $p$+Pb collisions with the trigger conditions
of the ALICE experiment \cite{ALICE_IpA} ($12<p_T^{\gamma}<40$ GeV),
the dominating contribution to the $pp$ and $pA$ photon-tagged FFs given by
(\ref{eq:30}) comes from the quark jets ($r_q/r_g\sim 20$).

As in \cite{Z_hl}, we calculate the medium-modified FFs $D_{h/i}^{m}$ 
using the triple $z$-convolution formula 
\beq
D_{h/i}^{m}(Q)\approx D_{h/j}(Q_{0})
\otimes D_{j/k}^{in}\otimes D_{k/i}^{DGLAP}(Q)\,,
\label{eq:40}
\eeq
where $D_{k/i}^{DGLAP}$ is the DGLAP FF for $i\to k$ transition,
$D_{j/k}^{in}$ is the in-medium $j\to k$ FF,
and  $D_{h/j}$ describes vacuum hadronization transition of
the parton $j$ to
hadron $h$.
We calculate the vacuum FFs $D_{h/i}^{v}(z,Q)$
using (\ref{eq:40}) with dropped $D_{j/k}^{in}$.
We use the KKP \cite{KKP} parametrization for
$D_{h/j}$  with $Q_0=2$ GeV.
We calculate the DGLAP FFs
$D_{k/i}^{DGLAP}$ using the PYTHIA event generator
\cite{PYTHIA}.
The medium dependence of $D_{h/i}^{m}$ given by (\ref{eq:40}) comes only
from the in-medium FFs $D_{j/k}^{in}$. We calculate them
from the one gluon spectrum in the approximation of the
independent gluon emission \cite{RAA_BDMS}.
As in \cite{Z_hl}, we account for collisional energy loss
(which is relatively small \cite{Z_coll}),
by treating it
as a perturbation to the radiative mechanism
with the help of a renormalization of the mQGP temperature in calculating
$D_{j/k}^{in}$.
As in \cite{Z_hl}, we calculate $D_{j/k}^{in}$ for 
an effective symmetrical  fireball with a uniform entropy/density distribution
in the transverse plane. We have checked that for a small size QGP this
approximation has a very good accuracy.
In calculating $D_{j/k}^{in}$, the averaging over the jet production points
(which corresponds to accounting for fluctuations of the parton path
length $L$ in the fireball) has been performed for the Gaussian parton
distribution in the transverse plane. However, we have found that the
$L$-fluctuations (and the shape of the distribution of the jet production
points in the transverse plane) are unimportant.
This occurs because for the expanding QGP the radiative energy loss 
to good accuracy $\propto L$ \cite{Z_pp13}. As a results, the predictions
for $I_{pA}$ turn out to be close to that obtained with the $pA$/$pp$
FFs $D_{h/i}^{m}$ for the average jet path lengths $\langle L\rangle$ (which
are close to the values of $R_f$).
We refer the interested reader for details of the numerical
calculations of $D_{h/i}^{m,v}$  to \cite{Z_hl}.

The induced gluon spectrum and the collisional energy loss have been
calculated with running $\alpha_s$. 
As in \cite{RAA20T,Z_hl}, we use the parametrization
(motivated by the lattice results of \cite{Bazavov_al1})
\beq
\alpha_s(Q,T) = \begin{cases}
\dfrac{4\pi}{9\log(\frac{Q^2}{\Lambda_{QCD}^2})}  & \mbox{if } Q > Q_{fr}(T)\;,\\
\alpha_{s}^{fr}(T) & \mbox{if }  Q_{fr}(T)\ge Q \ge cQ_{fr}(T)\;, \\
\frac{Q\alpha_{s}^{fr}(T)}{cQ_{fr}(T)} & \mbox{if }  Q < cQ_{fr}(T)\;, \\
\end{cases}
\label{eq:50}
\eeq
with $c=0.8$,
$Q_{fr}(T)=\Lambda_{QCD}\exp\left\lbrace
{2\pi}/{9\alpha_{s}^{fr}(T)}\right\rbrace$ (we take $\Lambda_{QCD}=200$ MeV)
and $Q_{fr}=\kappa T$.
We use $\kappa=2.55$, obtained in \cite{Z_hl} for scenario with the mQGP
formation in $pp$ collisions
by fitting of the LHC data on $R_{AA}$ for 2.76 and 5.02 TeV
Pb+Pb, and 5.44 TeV Xe+Xe collisions.\\

{\bf Model of the mQGP fireball in $pp$ and $pA$ collisions}.
We assume that in the midrapidity region the QGP evolution may be
described by Bjorken's model \cite{Bjorken} with 1+1D
isentropic longitudinal expansion. This gives the QGP entropy density
$s=s_0(\tau_0/\tau)$ at $\tau>\tau_0$
($\tau_0$ is the QGP formation proper time,
as in \cite{Z_hl}, we take $\tau_0=0.5$ fm), and use
a linear parametrization $s=s_0(\tau/\tau_0)$ for $\tau<\tau_0$.

We perform calculations of jet modification in $pp$/$pA$ collisions for symmetric fireballs.
This seems to be reasonable, since the azimuthal asymmetry is irrelevant for
the azimuthally averaged FFs that we need.
In Bjorken's model \cite{Bjorken},
with isentropic evolution of the fireball,
the initial entropy
density $s_0$ can be expressed as
\beq
s_{0}=\frac{C}{\tau_{0}\pi R_{f}^{2}}\frac{dN_{ch}^{f}}{d\eta}\,,
\label{eq:60}
\eeq
where $R_{f}$ is the fireball radius, $dN_{ch}^f/d\eta$ is the charged
hadron multiplicity pseudorapidity density
generated after hadronization of the QGP fireball, and
$C=dS/dy{\Big/}dN_{ch}/d\eta\approx 7.67$ is the entropy/multiplicity
ratio \cite{BM-entropy}.
We assume that in $pp$ collisions  the whole multiparticle production
 goes
 through hadronization of the isentropically expanding mQGP fireball,
 and consequently $dN_{ch}^f(pp)/d\eta=dN_{ch}^{ue}(pp)/d\eta$.
This seems to be reasonable, since for $pp$ collisions
the initial entropy deposition distribution should be
more or less uniform due to a small size of the interaction region
(of the order of the proton size, since $pp$ jet events are dominated
by nearly central $pp$ collisions).
By interpolating the ATLAS data \cite{ATLAS_UE_Nch} for
$dN_{ch}^{ue}(pp)/d\eta$ at $\sqrt{s}=0.9$ and $7$ TeV
(assuming that $dN_{ch}^{ue}(pp)/d\eta\propto s^{\delta}$)
we obtain $dN_{ch}^{ue}(pp)/d\eta\approx 12.5$ for $5.02$ TeV $pp$ collisions.
Using,  as in \cite{Z_hl,Z_Ipp}, the predictions 
for the multiplicity dependence of $R_f(pp)$ obtained
in the Color Glass Condensate (CGC) model \cite{glasma_pp,RPP},
we obtain $R_f(pp)\approx 1.49$ fm. This leads to the initial
fireball temperature $T_0(pp)\approx 225 $ MeV for the ideal
gas model entropy density,
and $T_0(pp)\approx 256$ MeV for the lattice entropy density \cite{t-lat}.

One can expect that for $pA$ jet events
$dN_{ch}^{f}(pA)/d\eta$  
should be somewhat smaller than the experimentally observed UE multiplicity
density $dN_{ch}^{ue}(pA)/d\eta$.
Indeed, in $pA$ collisions the typical UE in jet production 
includes one hard $pN$ interaction with
jet production (as for $pp$ collisions it should be dominated by
nearly central $pN$ collisions)  and several additional soft interactions
with ``spectator'' nucleons that are not involved in the jet production.
To understand the relative contribution to $dN_{ch}^{ue}(pA)/d\eta$
in $pA$ jet events of hadrons that are not related to the mQGP fireball,
we have performed simulation of the entropy deposition for $pA$ jet events
within the Monte Carlo wounded nucleon Glauber model
\cite{WNG,KN,PHOBOS_MC,GLISS2}. We used the form of the Monte Carlo
Glauber model suggested in \cite{MCGL1}. In Refs.~\cite{MCGL2,MCGL3},
this model was successfully used
 for description of a large amount of experimental data
on $AA$ and $pA$ collisions from RHIC and the LHC.

In the wounded nucleon Glauber model, we have
for the average midrapidity multiplicity density
in $pA$ minimum-bias events
(we use the form without the binary collision term,
since it gives the best fit to the experimental midrapidity multiplicity
in 5.02 TeV $p$+Pb collisions)
\beq
\frac{dN_{ch}^{mb}(pA)}{d\eta}=
\frac{dN_{ch}^{mb}(pp)}{d\eta}+
\frac{(N_w^{A}-1)}{2}
\frac{dN_{ch}^{mb}(pp)}{d\eta}\,,
\label{eq:70}
\eeq
where $dN_{ch}^{mb}(pp)/d\eta$ is the $pp$ minimum-bias multiplicity density
(as usual \cite{WNG}, the contribution of each wounded nucleon
equals $(1/2)dN_{ch}^{mb}(pp)/d\eta$),
and $N_w^{A}$ is the number of the wounded nucleons in the nucleus.
Our Monte Carlo simulation gives $N_w^A\approx 5.64$ for the minimum-bias
5.02 TeV $p$+Pb collisions.
With $dN_{ch}^{mb}(pp)/d\eta\approx 5.32$
for $5.02$ TeV $pp$ collisions
(obtained with the help of the power law interpolation
of the ALICE data \cite{ALICE_nch541} on
the charged multiplicity in NSD $pp$ events
at $\sqrt{s}=2.76$ and $7$ TeV) 
formula (\ref{eq:70}) gives $dN_{ch}^{mb}(pPb)/d\eta\approx 17.7$,
which agrees well
with the experimental value $dN_{ch}^{mb}(pPb)/d\eta\approx 17.8$
from the ALICE measurement \cite{ALICE_pA502}.

The UEs for jet events differ from the minimum-bias $pA$ collisions,
since for each UE we always have (at least) one hard $pN$ interaction,
which gives the multiplicity density $dN_{ch}^{ue}(pp)/d\eta$
(instead of the first term
$dN_{ch}^{mb}(pp)/d\eta$ on the right hand side of (\ref{eq:70})
for minimum-bias $pA$ collisions).
Then, it is natural to write the generalization of (\ref{eq:70}) to
the UEs in $pA$ collisions with jet production as
\beq
\frac{dN_{ch}^{ue}(pA)}{d\eta}=
\frac{dN_{ch}^{ue}(pp)}{d\eta}+
\frac{(N_w^{A}-1)}{2}
\frac{dN_{ch}^{mb}(pp)}{d\eta}\,.
\label{eq:80}
\eeq
For jet events $N_w^A$ is larger than for the minimum-bias events,
since jet events are biased to more-central $pA$ collisions.
Our Monte Carlo Glauber simulation of jet events in
$5.02$ TeV $p$+Pb collisions
gives $N_w^A\approx 9$.
With this value of $N_w^A$, (\ref{eq:80}) gives
$dN_{ch}^{ue}(pPb)/d\eta \approx 34.3$, which agrees
well with the average
UE charged multiplicity density for jet events found by ALICE \cite{ALICE_IpA}
in $5.02$ TeV $p$+Pb collisions.
The Monte Carlo simulation shows that in the $b$-plane the fireball has a
well pronounced peak at $r\lsim 1$ fm
(due to the UE multiplicity for the hard $pN$ collision with jet production,
which also leads to
the $dN_{ch}^{ue}(pp)/d\eta$ in (\ref{eq:80})),
and a broad corona
region at $r\gsim 1-1.5$ fm
formed by the spectator wounded nucleons (each of them
gives the multiplicity $0.5dN_{ch}^{mb}(pp)/d\eta \sim 2.65$).
At $r\sim 1.5-2$ fm the ideal gas QGP temperature
falls to $\sim 130-200$  MeV
(and falls steeply with rising $r$).
The entropy/multiplicity density in the corona region is close to or smaller
than that for $pp$ minimum-bias events at
$\sqrt{s}\sim 0.2$ TeV  ($dN_{ch}^{mb}(pp)/d\eta\sim 2.65$ \cite{UA1_pp}),
for which the probability of the QGP formation is expected to be small
\cite{ALICE_strange,Camp1}.
For this reason, it is reasonable to assume that,
only the core region is occupied by the mQGP fireball, and 
the corona wounded nucleons
produce hadrons in a nearly free-streaming regime.
Note that excluding the region with
the energy density corresponding to $T\lsim 130-200$ MeV from
the mQGP fireball is similar to the prescription of \cite{glasma_pp} used
for calculation of the mQGP fireball size within the CGC model.
To exclude the corona contribution to the mQGP fireball entropy we
 write the charged hadron multiplicity
associated with the mQGP fireball hadronization as
\beq
\frac{dN_{ch}^f(pA)}{d\eta}=
\frac{dN_{ch}^{ue}(pp)}{d\eta}+
\xi \frac{(N_w^{A}-1)}{2}
\frac{dN_{ch}^{mb}(pp)}{d\eta}\,.
\label{eq:90}
\eeq
Our Monte Carlo simulation shows that the number of the corona nucleons
may be as large as $\sim 0.5(N_w^A-1)$ (i.e. $\xi\sim 0.5$
in (\ref{eq:90})). For $\xi=0.5$ formula (\ref{eq:90}) gives
$dN_{ch}^{f}(pPb)/d\eta\approx 23.4$. Of course, this value of
$dN_{ch}^{f}(pPb)/d\eta$ is only a rough estimate.
Nevertheless, there is no reason to doubt that
a sizeable fraction of the UE hadron multiplicity in 5.02 TeV $p$+Pb collisions
may not be related to the mQGP hadronization.
Since the dynamics of the
non mQGP hadrons should be close to the free-streaming regime,
their effect on jet quenching should be small.

Our model neglects the size and density fluctuations
for the mQGP fireballs produced in $pp$ and $pA$ collisions.
In \cite{Z_Ipp} it has been argued  that for a small size QGP this
approximation is quite reasonable, since
due to the dominance of the $N=1$ rescattering contribution to induced
gluon emission, which has approximately linear dependence
on $L$ and density, the effect of the fireball size and density fluctuations
should be small.\\

{\bf Numerical results}.
In the absence of accurate calculations of the mQGP fireball
parameters for $pA$ collisions, we perform numerical
calculations of $I_{pA}$ for several values of $dN_{ch}^{f}(pA)/d\eta$
between the $pp$ and $pA$ UE charged multiplicity density
corresponding to $\xi=0,1/3,2/3$, and $1$ in (\ref{eq:90}).
 This set of $\xi$ leads to
$dN_{ch}^{f}(pA)/d\eta\approx12.5,19.8,27.1$, and $34.3$. 
We determine $R_f$ for $pA$ collisions using
the multiplicity dependence of $R_f$ obtained 
in the CGC numerical simulations performed in \cite{glasma_pp}. 
For our set of values of $\xi$/$dN_{ch}^{f}(pA)/d\eta$ we obtain
\beq
R_{f}(pA)[\xi=0,1/3,2/3,1]
\approx[1.62,1.85,2.03,2.16]\,\,\mbox{fm}\,.
\label{eq:100}
\eeq
Then, using the Bjorken relation (\ref{eq:60}),
we obtain for the initial temperature
defined via the ideal gas entropy and via the lattice entropy \cite{t-lat}
(numbers in brackets)
\beq
T_{0}(pA)[\xi=0,1/3,2/3,1]
\approx[214(244),228(257),237(267),246(275)]\,\,\mbox{MeV}\,.
\label{eq:110}
\eeq

\begin{figure} 
\begin{center}
\includegraphics[height=8.1cm,angle=-90]{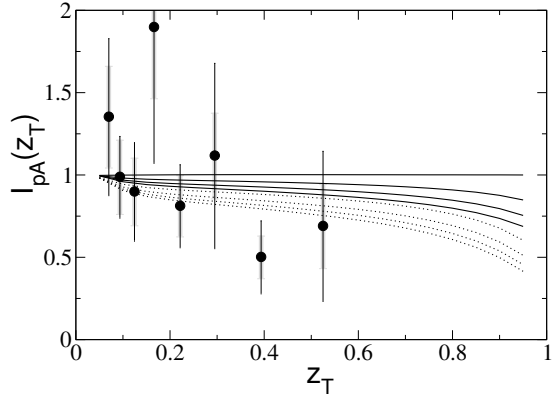}  
\end{center}
\caption[.]
        {The medium modification factor $I_{pA}$
of the photon-tagged FFs
          for $5.02$ TeV $p$+Pb
          collisions
          vs $z_T=p_T^{h}/p_T^{\gamma}$ for the trigger photon
          momentum window $12<p_T^{\gamma}<40$ GeV. Lines show the results
          of our calculations
with (solid) and without (dotted) the mQGP formation
          in $pp$ collisions
          for (top to bottom) $\xi=0,1/3,2/3,$ and 1 in the formula
          (\ref{eq:90}). 
Data points are from ALICE \cite{ALICE_IpA}.
}
\end{figure}

In Fig. 1 we plot the $z_T$-dependence of $I_{pA}$
for $\xi=0,1/3,2/3$, and $1$.
To illustrate the effect of jet quenching in $pp$ collisions
on $I_{pA}$, in Fig. 1 we show the results both for the scenarios
with (solid) and without (dotted) the mQGP formation in $pp$ collisions.
The area between the solid lines for $\xi=1/3$ and $2/3$ can be thought
as a reasonable theoretical uncertainty band for $I_{pA}$ 
in the scenario with the mQGP formation in $pA$ and $pp$ collisions
due to uncertainties in the corona contribution
to the UE multiplicity in $p$+Pb collisions.
The solid curve for $\xi=0$ in Fig.~1 corresponds to the mQGP entropy
in $p$+Pb collisions the same as that in $pp$ collisions. The equality
of the $pA$ and $pp$ fireball entropies results
in the same degree of medium suppression for the $pA$ and $pp$
photon-tagged FFs. For this reason, for $\xi=0$ we have $I_{pA}\approx 1$
(the effect of the difference in the nuclear PDFs and the proton PDFs,
that can affect the weight factors $r_i$ in (\ref{eq:30})
and lead to a deviation of $I_{pA}$ from unity,
turns out to be negligible).
As can be seen from
Fig. 1, $I_{pA}$ decreases with $z_T$. For $\xi\sim 0.5$ we have
$(1-I_{pA})\sim 0.1(0.2)$ at $z_T\sim 0.5$
and $(1-I_{pA})\sim 0.2(0.45)$ at $z_T\sim 0.9$
for scenario with(without) the mQGP formation
in $pp$ collisions.
The ALICE data \cite{ALICE_IpA} also show the tendency
of $I_{pA}$ to decrease with increasing $z_T$. However, the experimental errors
are large ($\sim \pm 0.5$ at $z_T\sim 0.5$), this fact does not allow to
validate or rule out the
scenario with the mQGP formation.
From the results shown in Fig. 1 one can conclude
that the use of the $\gamma$+jet process as a probe for jet quenching
in $pA$ and $pp$ collisions requires high accuracy data on $I_{pA}$
(with errors $\lsim 0.1-0.2$).

From the results presented in Fig.1 one can see that
for the scenario without the mQGP formation in $pp$ collisions
$(1-I_{pA})$ is larger than that for the scenario with
the mQGP formation both in $pA$ and $pp$ collisions by a factor of $2$.
This says that the effect of the medium modification of
the reference $pp$ FF $D_h^{pp}$ in the denominator of (\ref{eq:10})
is very important.
By itself the scenario with the mQGP formation only in $pA$ collisions
seems to be unrealistic,
since this scenario
is clearly inconsistent with data on the nuclear modification factor
$R_{pA}$ \cite{Z_RpA}.

To test the stability of the results with respect to variation
of $\tau_0$, we also performed calculations for $\tau_0=0.8$ fm.
Calculations with the same $\alpha_s$ (i.e. for $\kappa=2.55$) as
for $\tau_0=0.5$ fm, show very small variation of $I_{pA}$:
e.g. for $z_T\sim 0.5$ the value of $(1-I_{pA})$ is suppressed by
$\sim 1(5)$\% for version with(without) the mQGP formation in $pp$
collisions for $\xi=1$ (for which the changes in $I_{pA}$ are maximal).
Calculations with $\kappa\approx 2.43$, which corresponds to the
optimal $\chi^2$ fit of the LHC heavy ion data on $R_{AA}$
for $\tau_0=0.8$ fm \cite{Z_hl},
lead to nearly the same results for $I_{pA}$ as shown in Fig. 1.
Note also that for a given $dN_{ch}^f(pA)/d\eta$ we found
very little sensitivity
of $I_{pA}$ to the fireball radius $R_f$ (which we determined from the
IP-Glasma model calculations \cite{glasma_pp,RPP}).
This is due to a compensation of the
variations of parton energy loss arising from the
increase/decrease of the parton path length and from the decrease/increase
of the mQGP density. The low sensitivity of the jet quenching effects
to the mQGP fireball size was previously found for $R_{pp}$ \cite{Z_hl}.\\

{\bf Summary}.
We have calculated the medium modification factor $I_{pA}$
for the photon-tagged jets in 5.02 TeV $p$+Pb collisions
for the conditions of the ALICE experiment \cite{ALICE_IpA}
in the scenario with the mQGP formation.
Radiative and collisional
energy losses of fast partons in the QGP 
have been evaluated with running $\alpha_s(Q,T)$
that has a plateau around $Q\sim \kappa T$.
We perform calculations using $\kappa=2.55$ fitted to
the LHC heavy ion data on the nuclear modification factor $R_{AA}$.
Our calculations show that jet quenching can lead
to a deviation of $I_{pA}$ from unity by $\sim 0.1-0.2$ for
$z_T\sim 0.5-0.8$ for the scenario with the mQGP formation both in
$p$+Pb and $pp$ collisions. This, within errors, is consistent with
the data from ALICE \cite{ALICE_IpA}.
However, a definite conclusion about the presence or absence of jet quenching
in $pA$ collisions
cannot be drawn due to large experimental errors of the ALICE
data \cite{ALICE_IpA}.
Our results demonstrate that this requires a significantly more accurate
measurement of $I_{pA}$ (with errors $\lsim 0.1-0.2$).\\


\begin{acknowledgments}

  This work is supported by Russian Science Foundation
  grant  No. 20-12-00200
  in association with Steklov Mathematical Institute.\\

\end{acknowledgments}

\end{document}